\documentclass[preprint]{elsarticle}
\usepackage[utf8]{inputenc}
\usepackage{amssymb}
\usepackage{amsmath}
\usepackage[normalem]{ulem}
\usepackage[T1]{fontenc}
\usepackage{url}
\usepackage{hyperref}
\hypersetup{
    colorlinks=true,
    linkcolor=blue,
    filecolor=magenta,      
    urlcolor=blue,
    citecolor=blue,
}
\usepackage{dcolumn}
\usepackage{bm}
\usepackage{natbib}
\usepackage{float}
\usepackage{ulem}
\usepackage{color}
\usepackage{geometry}
\usepackage[caption=false,listofformat=subsimple,labelformat=simple]{subfig}

\usepackage{graphicx}
\setcitestyle{square,numbers}
\journal{Physica A}

\begin{document}

\begin{frontmatter}

\title{Blume-Emery-Griffiths model on Random Graphs}

\author{R. Erichsen Jr.} \ead{rubem@if.ufrgs.br}
\author{Alexandre Silveira} \ead{huyarius@gmail.com}
\author{S. G. Magalhães} \ead{sgmagal@gmail.com}

\address{Instituto de F\'{\i}sica, Universidade Federal do Rio Grande
  do Sul, Cx. Postal 15051, 91501-970 Porto Alegre, RS, Brazil}

\begin{abstract}
The Blume-Emery-Griffiths model with a random crystal field is studied
in a random graph architecture, in which the average connectivity is a
controllable parameter. The disordered average over the graph
realizations is treated by replica symmetry formalism of order
parameter functions. A self consistent equation for the distribution
of local fields is derived, and numerically solved by a population
dynamics algorithm. The results show that the average connectivity
amounts to changes in the topology of the phase diagrams.  Phase
diagrams for representative values of the model parameters are
compared with those obtained for fully connected mean field and
renormalization group approaches.
\end{abstract}


\end{frontmatter}

\section{\label{sec:intro}Introduction}

The three spin states ($\sigma=0$, $\pm 1$) Blume-Emery-Griffiths
(BEG) \cite{BEG} was introduced with the aim to describe qualitatively
superfluidity in $^{3}$He - $^{4}$He mixtures and phase separation. It
is composed by three terms: the spin exchange interaction responsible
by stabilizing a magnetic order, the local crystal field favoring
non-active spin states $\sigma=0$ and the non-local bi-quadratic
interaction term favouring active spin state $\sigma=\pm 1$ in
neighboring sites. The competition between these three mechanisms is
responsible for giving rise to a complex phase diagram. For instance,
it is expected as an outcome, a phase diagram with second and
first-order phase transitions lines and multicritical points. This has
motivated this model to be investigated using several methods, such as
mean field theory \cite{PhysRevLett.67.1027}, effective field theory
\cite{Tucker1989}, cluster variation method \cite{Grigelionis1989,
  PhysRevB.47.2643}, Monte Carlo simulations
\cite{wang1,wang2,Kasono92}, Bethe lattice
\cite{Akheyan_1996,CHAKRABORTY1986122,Osorio_1989} and renormalization
group with hierarchical lattices \cite{SNOWMAN20093007}. This interest
in the BEG model raises the question of what might be the effects of
disorder on it.

It is known that the presence of disorder might lead to changes in the
boundary line order of the phase transitions and, consequently,
affecting multicritical phase diagrams \cite{PhysRevLett.35.1399,
  PhysRevLett.62.2507}. In the case of the BEG model, disorder can be
introduced in three ways: by choosing exchange interaction, crystal
field or bi-quadratic exchange strengths as a random variable or even
a combination of the previously mentioned possibilities. Each
situation can describe different problems.  For instance, the case of
a random bi-quadratic exchange can be used in neural networks, where
this term in the BEG model becomes a learning rule
\cite{PhysRevE.68.062901, 2005EPJB...47..281B}. On the other hand, a
random crystal field can be applied to the modeling mixtures
$^{3}$He-$^{4}$He in porous medium such as aerogel
\cite{PhysRevLett.74.426,PhysRevLett.71.2268}.
 
The three possibilities of disorder in the BEG model and its
combinations have been treated also in several techniques, such as
mean field, renormalization group, Bethe lattice, transfer matrix,
cluster variation, effective field theory (see, for instance,
Refs. \cite{PhysRevLett.62.2507, arenzon, ALBAYRAK2015107,
  Albayrak2015, DONG200790, Kple2021, 2020IJTP...59.3915K,
  KARIMOU20172371, Dong2009, Dong2007, Dong2006, PhysRevB.60.1033,
  Buzano1994, PhysRevLett.69.221}).  In the case of the random crystal
field field displayed in Ref. \cite{PhysRevB.60.1033}, results coming
from mean field approximation (i. e., with infinite dimension or
coordination number) and real space renormalization group can be
compared. This last technique is quite suitable to describe low
dimensionality scenarios.  The most important difference between the
two techniques is the suppression of the first-order phase transition
lines or their replacing by continuous ones obtained in the
renormalization group. The random crystal field BEG model with
anti-ferromagnetic (AF) bi-quadratic coupling constant in the Bethe
lattice was investigated in ref. \cite{Kple2021}.

Our purpose with this work is to study the random crystal field BEG
(RCBEG) model on the ensemble of poissonian random graphs. The random
graph offers the average connectivity $c$ as a continuous,
controllable parameter, allowing to investigate the RCBEG model for
different regimes, i.e., from large connectivity, close to the fully
connected limit corresponding to the mean field approximation, till to
the small connectivity situation. One can expect that might occur
important changes as compared with the mean field results involving
the replacement and/or disappearance of multicritical points in the
phase diagrams as long as $c$ decreases. Indeed, this kind of changes
for small $c$ has been confirmed in the Blume-Capel model
\cite{Blume,CAPEL1966966} with an added disorder given by a random
field. In that model, it was found that variations in $c$ produced
drastic changes in the multicritical phase diagrams as compared with
fully connected case \cite{Kaufman1990}. Indeed, some multicritical
points disappear when $c$ decreases \cite{rubemBC}.

To sum over the realizations of the random graph, we use the replica
symmetry theory of order parameter functions
\cite{monasson,Monasson_1998,Wemmenhove_2003}. As frustration is
absent is this model, we anticipate that the replica symmetry solution
is exact for this purpose. The same equations for this problem can be
derived by the cavity method \cite{cavitymp} after taking the ensemble
average \cite{lupo}. We approach the problem considering that the
lattice of spins is a random graph, where the connectivity is finite
and the degree of a site is given by a Poisson distribution. Thus, we
offer an alternative route to approach this problem. Also, we study
simultaneously the presence of random crystal field disorder and the
disorder of the lattice. As it has been shown, the connectivity has a
crucial role in phase diagram topology
\cite{rubemBC,RubemWalter,PhysRevE.103.022133}, allowing to change the
nature of transitions and critical points through a fine-tuning of the
control parameter

The paper is organized as follows. In Sec. \ref{sec:method} we
describe our model and derive the fundamental equations using replica
symmetry theory for finite connectivity systems. In
Sec. \ref{sec:results} we explain the method to numerically calculate
the distribution of fields, some examples of order parameters are
shown and the behavior of the system is described by drawing phase
diagrams for the thermodynamic phases. The conclusions can be found in
Sec. \ref{sec:ccl}.

\section{\label{sec:method} Model and Replica Procedure}

The model's Hamiltonian is
\begin{align}  
H(\boldsymbol{\sigma})=-\frac{J}{c}\sum_{i< j}c_{ij}
\sigma_{i}\sigma_{j} - \frac{K}{c}\sum_{i< j}c_{ij}
\sigma_{i}^{2}\sigma_{j}^{2}+\sum_{i}\Delta_{i}\sigma_{i}^{2}\,,
\label{hamiltonian1}
\end{align}
where $\mathcal{\sigma}\equiv\{\sigma_i\}\,,i=1\dots N $ denotes the
state of the system, $c_{ij}$ are independent, identically distributed
random variables (i.i.d.r.v.) chosen from the distribution
\begin{align}  
p\left(c_{ij}\right)=\frac{c}{N}\delta(c_{ij}-1) +
\left(1-\frac{c}{N}\right)\delta(c_{ij})\,,
\label{cdistr}
\end{align}
indicating if the pair of spins $i$ and $j$ is connected ($c_{ij}=1$)
or not ($c_{ij}=0$), with the constant $c$ representing the mean
connectivity. The local, random crystal fields $\Delta_i$ are
i.i.d.r.v. chosen from the distribution
\begin{align} 
p\left(\Delta_{i}\right)=p\delta(\Delta_{i}-\Delta)
+\left(1-p\right)\delta(\Delta_{i})\,.
  \label{Kdistr}
\end{align}
The constant $K$ controls the strenght of the bi-quadratic
couplings. Using the replica method we can write the disorder averaged
free energy as
\begin{align}
  f(\beta)=-\lim_{N\rightarrow\infty}\frac{1}{\beta
    N}\lim_{n\rightarrow 0} \frac{1}{n}\log\langle
  {Z^n}\rangle_{\mathbf{c},\boldsymbol{\Delta}}\,,
  \label{free}
\end{align}
where
\begin{equation}
Z^n=\sum_{\boldsymbol{\sigma}_{1}\dots\boldsymbol{\sigma}_{n}}
\mathrm{e}^{-\beta\sum_{\alpha} H(\boldsymbol{\sigma}_\alpha)}\,
\label{part}
\end{equation}
is the replicated partition function
$\boldsymbol{\sigma}_\alpha\,,\alpha=1\dots n$ denotes the state of
replica $\alpha$,
$\langle\cdot\rangle_{\mathbf{c}\boldsymbol{\Delta}}$, with
$\mathbf{c}\equiv\{c_{ij}\}$ and
$\boldsymbol{\Delta}\equiv\{\Delta_{i}\}$, denotes the disorder
average. In the limit $c/N\rightarrow 0$, the average over $c_{ij}$
gives
\begin{align}
\langle Z^{n} \rangle =
\sum_{\boldsymbol{\sigma}_{1}\dots\boldsymbol{\sigma}_{n}}
\langle\mathrm{e}^{-\beta\sum_{\alpha,i}\Delta_{i}
  \sigma_{i\alpha}^{2}}\rangle_{\boldsymbol{\Delta}}\exp\Big[\frac{c}{2N}\sum_{i\neq
    j}\Big(\mathrm{e}^{\frac{\beta J}{c}\sum_{\alpha}
    \sigma_{i\alpha}\sigma_{j\alpha}+\frac{\beta K}{c}\sum_{\alpha}
    \sigma_{i\alpha}^{2}\sigma_{j\alpha}^{2}}-1\Big)
  \Big]\,. \label{part1}
\end{align}
To transform into a single spin problem, order functions
\begin{equation}
P(\boldsymbol{\sigma})=\frac{1}{N}\sum_{i}
\delta_{\boldsymbol{\sigma}\boldsymbol{\sigma}_{i}}\,,
\end{equation}
which represent the probability of a replicated spin variable
$\boldsymbol{\sigma}_{i}$ to assume the replica state
$\boldsymbol{\sigma}$, and their conjugated order functions
$\hat{P}(\boldsymbol{\sigma})$, are introduced. The partition function
can be rewritten as (see the appendix)
\begin{align}
\langle Z^{n} \rangle=&\int\prod_{\boldsymbol{\sigma}}
d\hat{P}(\boldsymbol{\sigma})d P(\boldsymbol{\sigma})\exp
N\Big\{\sum_{\boldsymbol{\sigma}}\hat{P}(\boldsymbol{\sigma})
P(\boldsymbol{\sigma})+\frac{c}{2}\sum_{\boldsymbol{\sigma}\boldsymbol{\sigma}'}
P(\boldsymbol{\sigma})P(\boldsymbol{\sigma}')\nonumber\\ &
\times\Big(\mathrm{e}^{\frac{\beta J}{c}\sum_{\alpha}
  \sigma_{\alpha}\sigma_{\alpha}^{\prime}+\frac{\beta
    K}{c}\sum_{\alpha} \sigma_{\alpha}^{2}\sigma_{\alpha}^{\prime
    2}}-1\Big)+
\log\sum_{\boldsymbol{\sigma}}\langle\mathrm{e}^{-\hat{P}(\boldsymbol{\sigma})
  -
  \beta\Delta\sum_{\alpha}\sigma_{\alpha}^{2}}\rangle_{\Delta}\Big\}\,.  \label{RSsp}
\end{align}
In the thermodynamic limit the integral can be evaluated through the
saddle-point method. We eliminate the $\hat{P}(\boldsymbol{\sigma})$'s
through the saddle-point equations and rewrite the free-energy as
\begin{align}
\label{freenophat}
f(\beta)&=-\lim_{n\rightarrow 0} \frac{1}{\beta
  n}\mathrm{Extr}\Big\{-\frac{c}{2}\sum_{\boldsymbol{\sigma}
  \boldsymbol{\sigma}^{\prime}}
P(\boldsymbol{\sigma})P(\boldsymbol{\sigma}^{\prime})
\Big(\mathrm{e}^{\frac{\beta J}{c}\sum_{\alpha}
  \sigma_{\alpha}\sigma_{\alpha}^{\prime} + \frac{\beta
    K}{c}\sum_{\alpha} \sigma_{\alpha}^{2}\sigma_{\alpha}^{\prime
    2}}-1\Big)\\ & + \ln\Big\langle\sum_{\boldsymbol{\sigma}}
\exp{\Big[c\sum_{\boldsymbol{\sigma}^{\prime}}P(\boldsymbol{\sigma}^{\prime})
    \Big(\mathrm{e}^{\frac{\beta J}{c}\sum_{\alpha}
      \sigma_{\alpha}\sigma_{\alpha}^{\prime}+\frac{K
        \beta}{c}\sum_{\alpha}
      \sigma_{\alpha}^{2}\sigma_{\alpha}^{\prime 2}}-1\Big) -
    \beta\Delta\sum_{\alpha}\sigma_{\alpha}^{2}\Big]
  \Big\rangle_{\Delta}}\Big\}\,, \nonumber
\end{align}
where $\mathrm{Extr}$ amounts to take the extreme of the expression
between braces relatively to $P(\boldsymbol{\sigma})$, which gives the
remaining saddle-point equations
\begin{equation}
P(\boldsymbol{\sigma})=\frac{1}{\mathcal{N}}\Big\langle
\exp{\Big[c\sum_{\boldsymbol{\sigma}^{\prime}}
    P(\boldsymbol{\sigma}^{\prime})\Big(\mathrm{e}^{\frac{\beta J}{c}
      \sum_{\alpha}\sigma_{\alpha}\sigma_{\alpha}^{\prime} +
      \frac{K\beta}{c}\sum_{\alpha}\sigma_{\alpha}^{2}\sigma_{\alpha}^{\prime
        2}} - 1\Big) -
    \beta\Delta\sum_{\alpha}\sigma_{\alpha}^{2}\Big]\Big\rangle_{\Delta}}\,,
\label{RS1}
\end{equation}
where $\mathcal{N}$ is a normalization factor. 

We search solutions of Eq. (\ref{RS1}) satisfying the RS Ansatz, where
the order function is invariant under replica index permutations,
which are written in the form
\begin{equation}
P(\boldsymbol{\sigma}) =\int\mathcal{D}W(x{,}y)
\frac{\mathrm{e}^{\beta x\sum_{\alpha}\sigma_{\alpha} + \beta
    y\sum_{\alpha}\sigma^{2}_{\alpha}}}{\Big(\sum_{\sigma}\mathrm{e}^{\beta
    x\sigma + \beta y\sigma^{2}}\Big)^{n}}\,,
\label{RS}
\end{equation}
where $\mathcal{D}W(x{,}y)\equiv dxdyW(x{,}y)$. Expanding the
exponential of Eq. (\ref{RS1}) and introducing Eq. (\ref{RS}) we
obtain a self consistent equation for the distribution of local fields
(details in the Appendix)
\begin{align}
\label{fieldist}
W(x{,}y) =
\sum_{k=0}^{\infty}\frac{c^{k}\mathrm{e}^{-c}}{k!}\Big\langle\int
\prod_{l=1}^{k}\mathcal{D}W(x_l{,}y_l)
\delta\Big[x-\frac{1}{\beta}\sum_{l}\phi(x_l{,}y_l)\Big]\delta\Big[y+
  \Delta -
  \frac{1}{\beta}\sum_{l}\psi(x_l{,}y_l)\Big]\Big\rangle_{\Delta}\,,
\end{align}
where
\begin{equation}
\phi(x{,}y)=\frac{1}{2}\ln\frac{\chi_{+1}(x{,}y)}{\chi_{-1}(x{,}y)}\,,    
\end{equation}
\begin{equation}
\psi(x{,}y)=\frac{1}{2}\ln\frac{\chi_{+1}(x{,}y)\chi_{-1}(x{,}y)}
    {\chi_0^2(x{,}y)}\,,
\end{equation}
and
\begin{equation}
\chi_{\sigma}(x{,}y)=\sum_{\tau}\mathrm{e}^{\beta x\tau + \frac{\beta}{c}J
  \sigma\tau + \beta y\tau^{2} + \frac{\beta}{c}K
  \sigma^{2}\tau^{2}}\,.
\end{equation}

The relevant observables are the average magnetization
\begin{equation}
m=\sum_{\boldsymbol{\sigma}}\sigma_{\alpha}P(\boldsymbol{\sigma}) =
\int\mathcal{D}W(x,y)\frac{2\sinh(\beta x)}{\mathrm{e}^{-\beta y} +
  2\cosh(\beta x)}\,
\label{magnet}
\end{equation}
and the occupation number
\begin{equation}
Q=\sum_{\boldsymbol{\sigma}}\sigma_{\alpha}^{2}P(\boldsymbol{\sigma})
= \int\mathcal{D}W(x,y)\frac{2\cosh(\beta x)}{\mathrm{e}^{-\beta y} +
  2\cosh(\beta x)}\,.
\label{occ}
\end{equation}
To determine the RS free-energy we insert the Ansatz (\ref{RS}) in
Eq. (\ref{freenophat}) and take the limit $n\rightarrow 0$, which
results
\begin{align}
\nonumber f(\beta)=\frac{c}{2\beta} &\int \mathcal{D}W(x{,}y)
\mathcal{D}W(x'{,}y') \frac{\sum_{\sigma\sigma'} \mathrm{e}^{\beta
    x\sigma + \beta y\sigma^{2}+\beta x^{\prime}\sigma^{\prime} +
    \beta y^{\prime}\sigma^{\prime 2}+\frac{\beta}{c}J
    \sigma\sigma'+\frac{\beta}{c}K
    \sigma^2\sigma'^2}}{\chi_{0}(x{,}y)\chi_{0}(x'{,}y')}
\nonumber\\ & -\frac{1}{\beta}\sum_{k=0}^{\infty}P_k\int
\prod_{l=1}^{k}\mathcal{D}W(x_l{,}y_l)
\Big\langle\ln\Big(\sum_{\sigma}\mathrm{e}^{-\beta\Delta\sigma^{2}}
\prod_{l}\frac{\chi_{\sigma}(x_l{,}y_l)}{\chi_0(x_l{,}y_l)}\Big)
\Big\rangle_{\Delta}\,,
  \label{free1}
\end{align}
where $P_k=c^k\mathrm{e}^{-c}/k!$ is a poissonian weight.

\section{Results\label{sec:results}}

\begin{figure}
\centering
\includegraphics[width=8cm,clip]{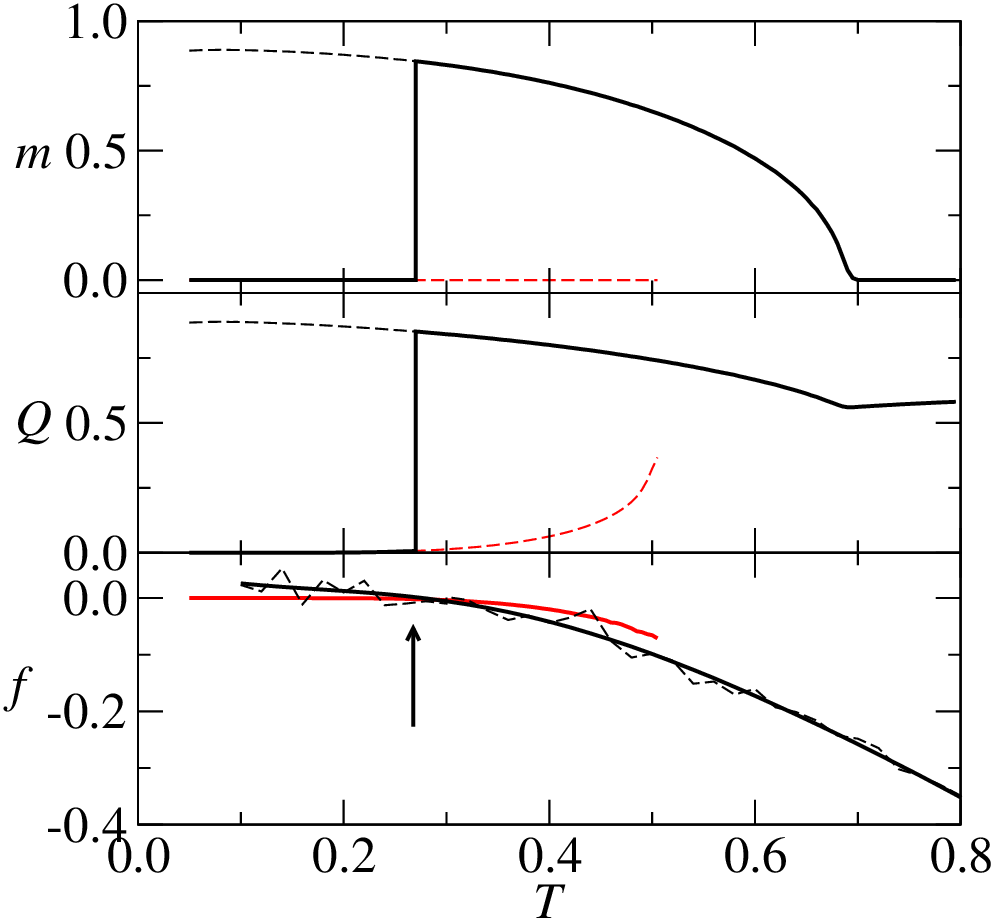}
\caption{ Magnetization $m$, occupation number $Q$ and free-energy as
  functions of $T$ for $p=1$, $c=8$, $K=2$ and $D = 1.55$. Solid black
  lines on $m$ and $Q$ represent the stable order parameters
  values. Dashed black (dashed red) line represents metastable FM (PM)
  solution. Dashed black line on $f$ is the metastable FM free-energy
  raw data. Solid black line is a polynomial adjust of the FM
  data. Solid red line is the PM free-energy data. The arrow signals
  the crossing of FM and PM free-energies.
  \label{fig:mQf_T_8_.05}}
\end{figure}

\begin{figure}
\centering
\includegraphics[width=8cm,clip]{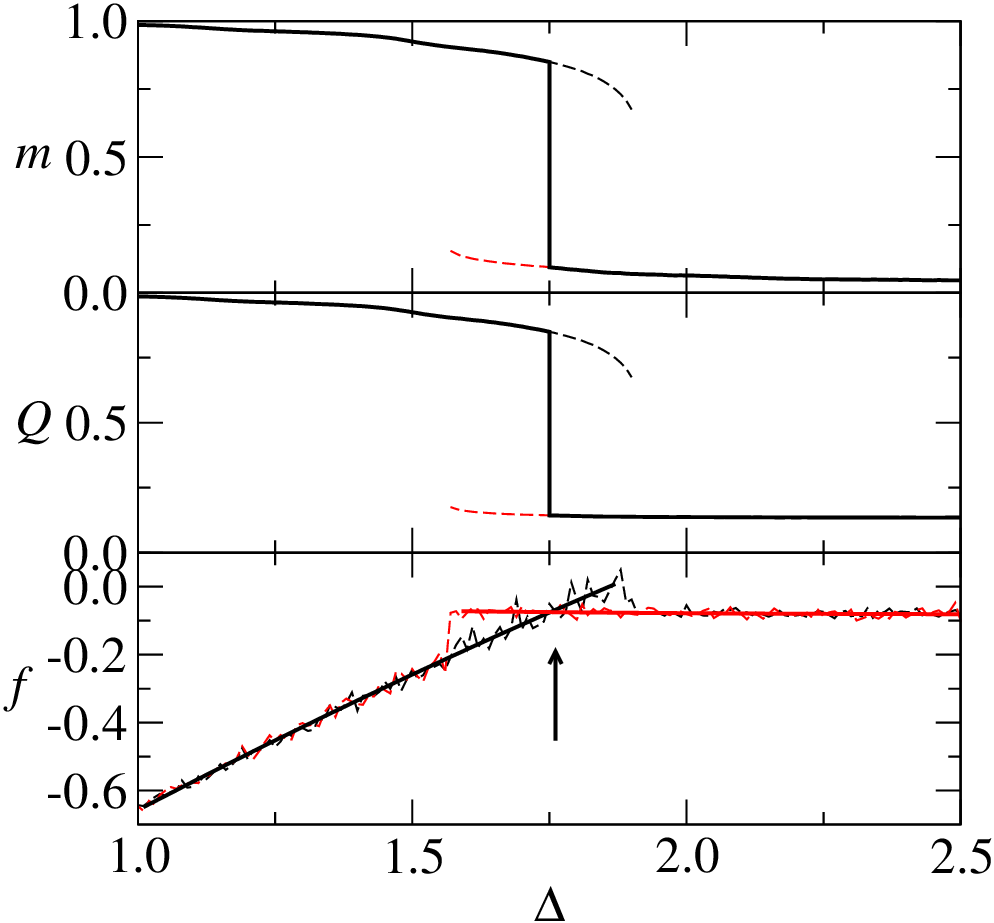}
\caption{ Magnetization $m$, occupation number $Q$ and free-energy as
  functions of $\Delta$ for $p=0.85$, $c=8$, $K=2$ and $T =
  0.05$. Solid black lines on $m$ and $Q$ represent the stable order
  parameters values. Dashed black (dashed red) line represents
  metastable FM (PM) solution. Dashed black (red) line on $f$ is the
  FM (PM) free-energy raw data. Solid black (red) line is a polynomial
  adjust of the FM (PM) data. The arrow signals the crossing of FM and
  PM free-energies.}
  \label{fig:mQf_D_8_.05}
\end{figure}

\begin{figure}
  \centering
  \includegraphics[width=8cm,clip]{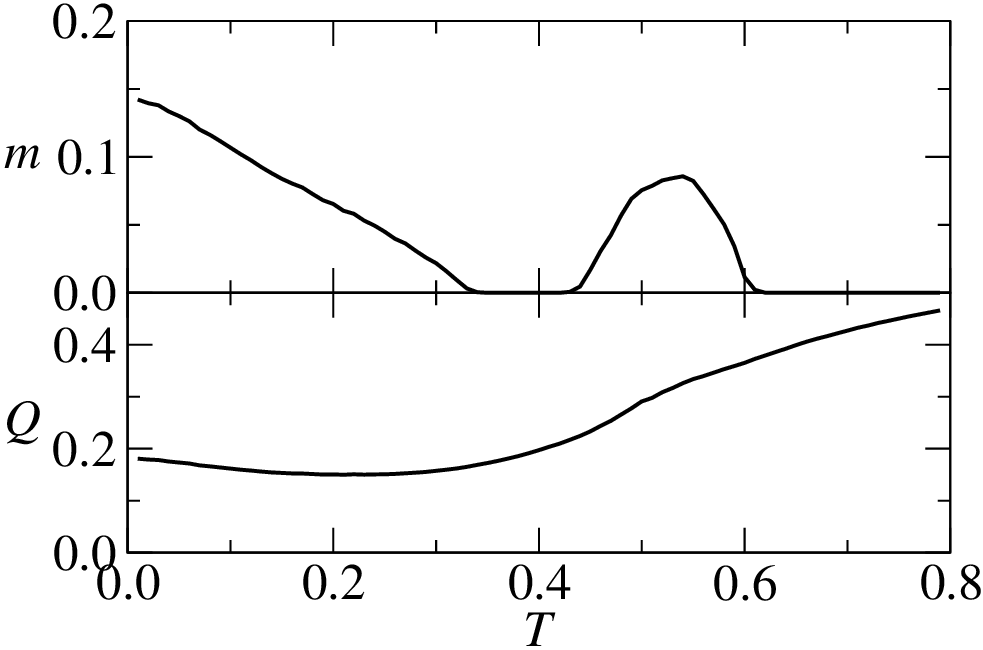}
  \caption{Magnetization $m$ and occupation number $Q$ as functions of
    $T$ for $p=0.85$, $c=4$, $K=2$ and $\Delta = 2.07$.}
\label{fig:mQ_2_.85}
\end{figure}

According to Eqs. (\ref{magnet}) -- (\ref{free1}), the relevant order
parameters are obtained through the calculation of the local field
distribution, given by the self consistent equation
(\ref{fieldist}). This is done numerically, via a population dynamics
algorithm \cite{cavitymp}, as follows: (i) a population of
$\mathcal{N}$ two-component fields $(x{,}y)$ is created; (ii) an
integer $k$ is randomly sorted from a Poisson distribution of mean
$c$, and $k$ fields are randomly chosen from the population; (iii)
with the sorted fields, evaluate the two summations appearing in the
delta functions of Eq. (\ref{fieldist}) and (iv) the results are
assigned to the components of a further randomly chosen field
$(x^{*}{,}y^{*})$. The algorithm is repeated till the convergence to a
stable population distribution $W(x,y)$. Throughout this work we used
populations of $\mathcal{N}=100{,}000$ fields and convergence time
that amounts to 5,000,000 iterations. Still, each point is averaged
over 20 runs. As shown in Eq. (\ref{free1}), the first free-energy
term contains a double integral and the second term contains a
$k$-fold integral over the local field distribution. To evaluate these
terms, we follow a Montecarlo algorithm: a large number (1,000,000) of
pairs and $k$-sets of local fields are randomly chosen and and their
contributions are summed. This results in a noisy curve, contrary to
$m$ and $Q$ evaluations that contain a simple integral. To overcome
the noise, the $f$ curves are adjusted by a polynomial.

\begin{figure}
    \centering
    \includegraphics[width=6.5cm,clip]{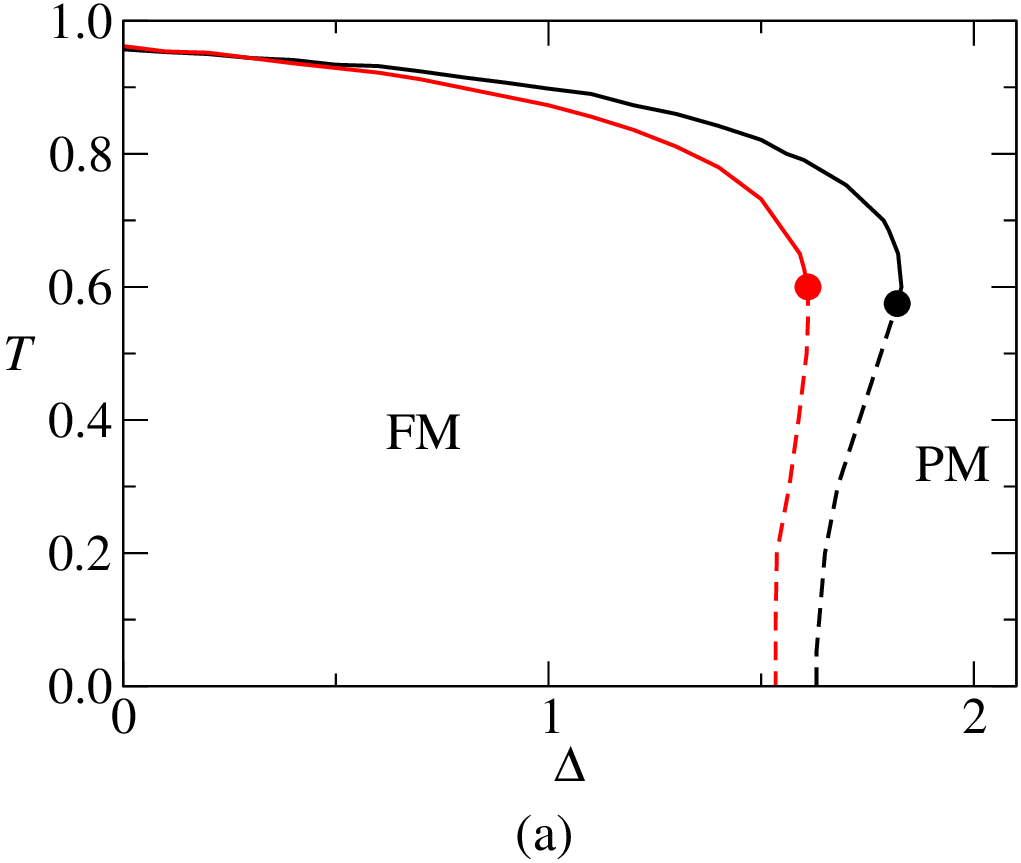}\hspace{0.5cm}
    \includegraphics[width=6.5cm,clip]{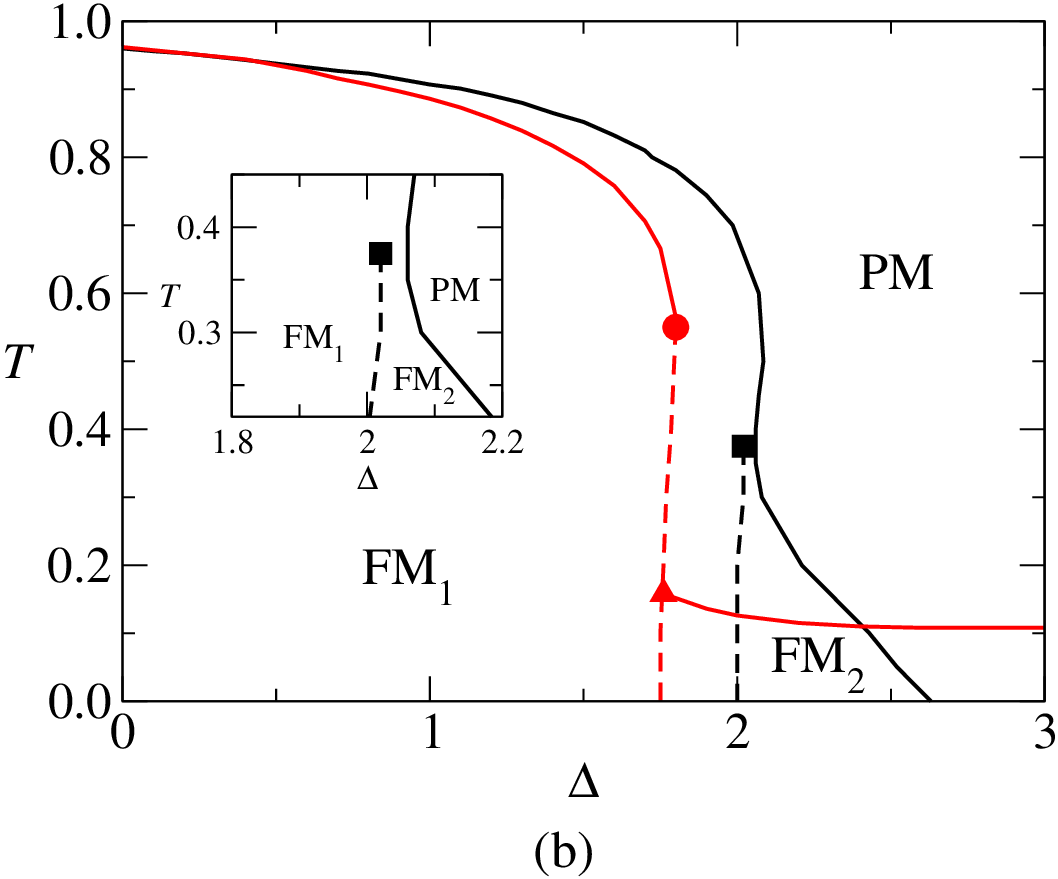}
    
    \vspace{0.5cm}
    \includegraphics[width=6.5cm,clip]{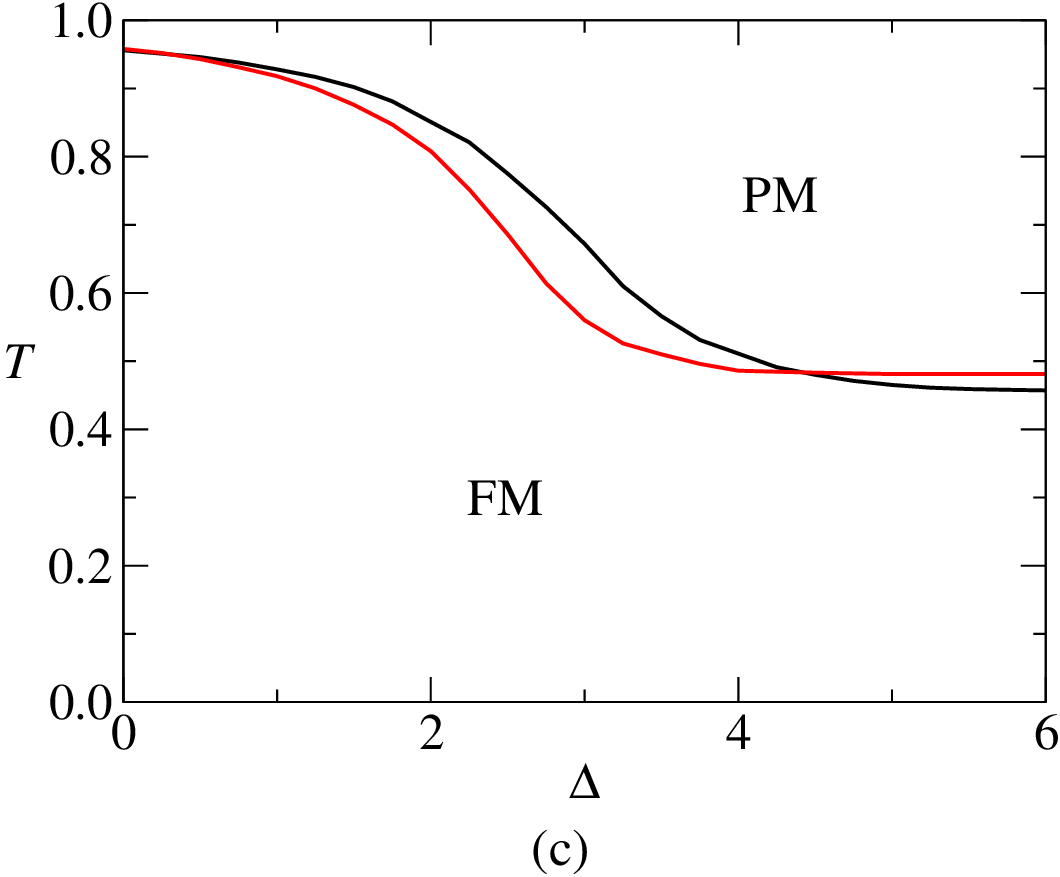}\hspace{0.5cm}
    \includegraphics[width=6.5cm,clip]{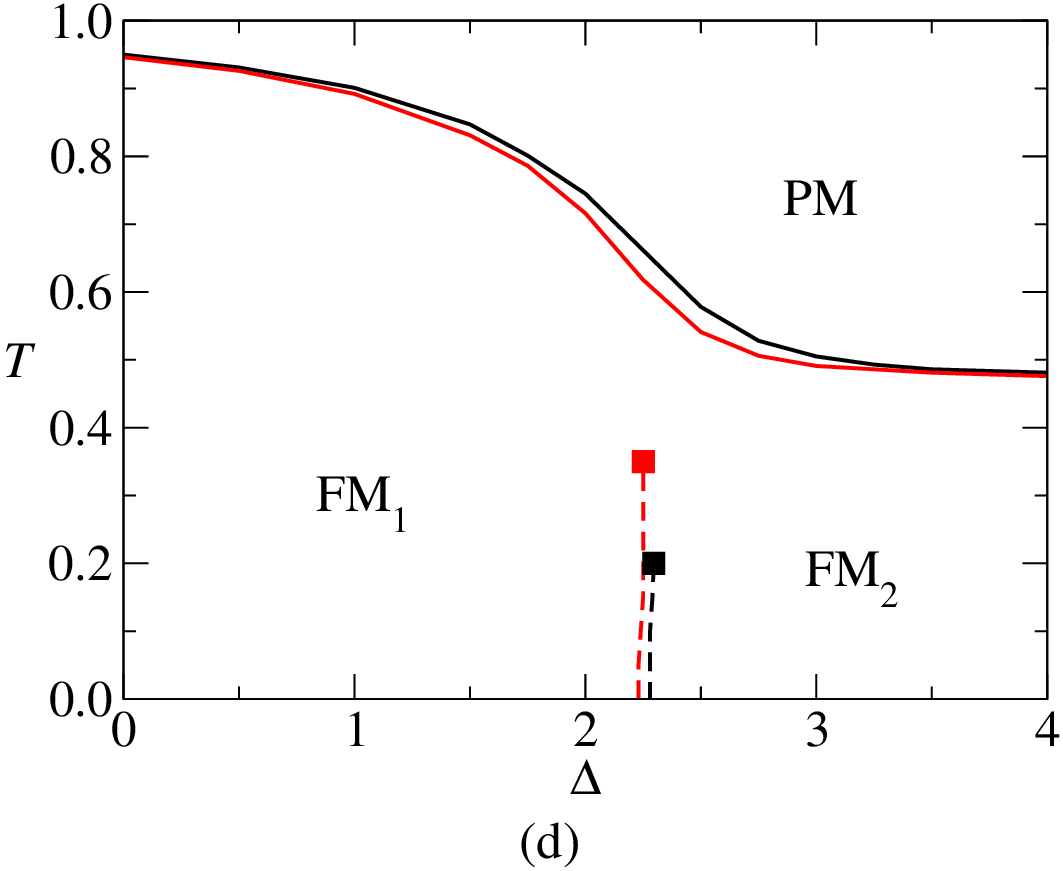}

    \caption{Thermodynamic phase diagrams $T$ versus $\Delta$ for
      $K=2$; (a) $p=1$, (b) $p=0.85$ and (c) $p=0.5$. The inset in (b)
      shows in detail the vicinity of the critical point. The
      connectivity values are $c=4$ (black) and $c=8$ (red). (d)
      Thermodynamic phase diagrams for $p=0.5$, $K=2$, $c=25$ (black),
      $c=100$ (red). Solid (dashed) lines correspond to continuous
      (discontinuous) transitions. Circles, squares and triangles
      represent tri-critical points, critical points and critical end
      points, respectively }
    \label{fig:TD_2}
\end{figure}

As example of the outcome, order parameters and free-energy curves are
shown in Fig. \ref{fig:mQf_T_8_.05}, as functions of $T$, for $c=8$,
$K=2$, $p=1$ and $\Delta=1.55$.  Here and in the sequel the energy
scale is fixed by assuming the bi-linear coupling constant $J=1$. For
$\Delta=1.55$, PM and FM phases coexist from $T=0$ till a continuous
FM -- PM transition at $T\approx 0.693$. To overcome the noisy
free-energy and find the discontinuous transition locus we resort to a
polynomial fit which indicates the crossing of the free-energy curves
at $T\approx 0.266$. PM is stable in the $0\leq T\lesssim 0.266$ and
$0.693\lesssim T$ interval. FM is stable in the $0.266\lesssim
T\lesssim 0.693$ interval. This characterizes a re-entrant
behavior. The discontinuous transition at $T\approx 0.266$ appears as
a dashed red line on Fig. \ref{fig:TD_2}a and continuous transition at
$T\approx 0.693$ appears as a solid red line in the same figure.

Order parameters and free-energy curves as functions of the crystal
field $\Delta$, for $c=8$, $K=2$, $p=0.85$ and $T=0.05$ are shown in
Fig. \ref{fig:mQf_D_8_.05}. The curves show a high $m$, high $Q$
FM$_{1}$ phase at small $\Delta$, a low $m$, low $Q$ FM$_{2}$ at large
$\Delta$ and a co-existence region between them. As mentioned above,
we resort to a linear fit to find a crossing of the free-energy curves
at $\Delta\approx 1.75$. This reveals a discontinuous transition
between the two ferromagnetic phases, represented by the dashed red
line on Fig. \ref{fig:TD_2}b.  The reason for the existence of two
FM's phases will be discussed below.

The order parameters $m$ and $Q$ as functions of the temperature for
$c=4$, $p=0.85$, $K=2$ and $\Delta=2.07$ are shown in
Fig. \ref{fig:mQ_2_.85}. This figure shows, as the temperature
increases, a FM$_2$ phase, then a re-entrant PM phase, a FM$_2$ phase
and a PM phase at high $T$.

\begin{figure}
\centering \includegraphics[width=8cm,clip]{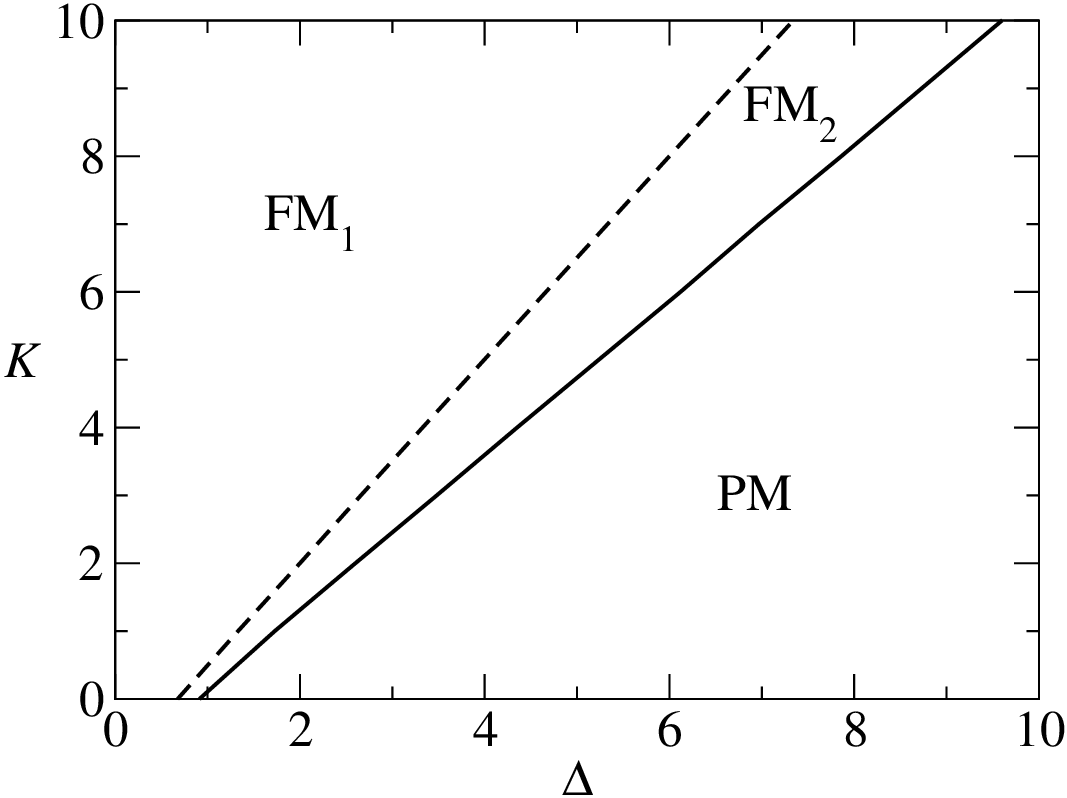}
\caption{$K$ versus $\Delta$ phase diagram for $T=0$, $c=4$ and
  $p=0.85$. Solid (dashed) lines corresponds to continuous
  (discontinuous) transition.}
\label{fig:KD}
\end{figure}

To give a complete overview of a model with so many parameters,
keeping a reasonable amount of pictures, is a difficult task, and the
zero-temperature $K$ versus $\Delta$ phase diagram may guide us. This
diagram is shown in Fig. \ref{fig:KD} for the representative case
$c=4$ and $p=0.85$, revealing a discontinuous FM$_1$ - FM$_2$
transition and a continuous FM$_2$ - PM transition. The two
ferromagnetic phases are present, at low temperature, whenever $p<1$,
i.e., in the presence of disorder. This disorder acts turning off the
crystal field $\Delta$ in a $1-p$ fraction of sites, this way
favouring the active states in these sites. The higher magnetization
FM$_1$ is found at low $\Delta$ value, while the lower magnetization
FM$_2$ and PM are found for higher $\Delta$'s. Since the bi-quadratic
coupling constant $K$ favours the active states, higher magnetization
phases are found as $K$ increases. It is unnecessary to add further
zero temperature diagrams, but it is worthy to mention that, as the
connectivity $c$ increases, or $p$ decreases, FM$_2$ becomes stable at
large $\Delta$ and there is no more a PM phase at $T=0$.

\begin{figure}
\centering
    \includegraphics[width=6.5cm,clip]{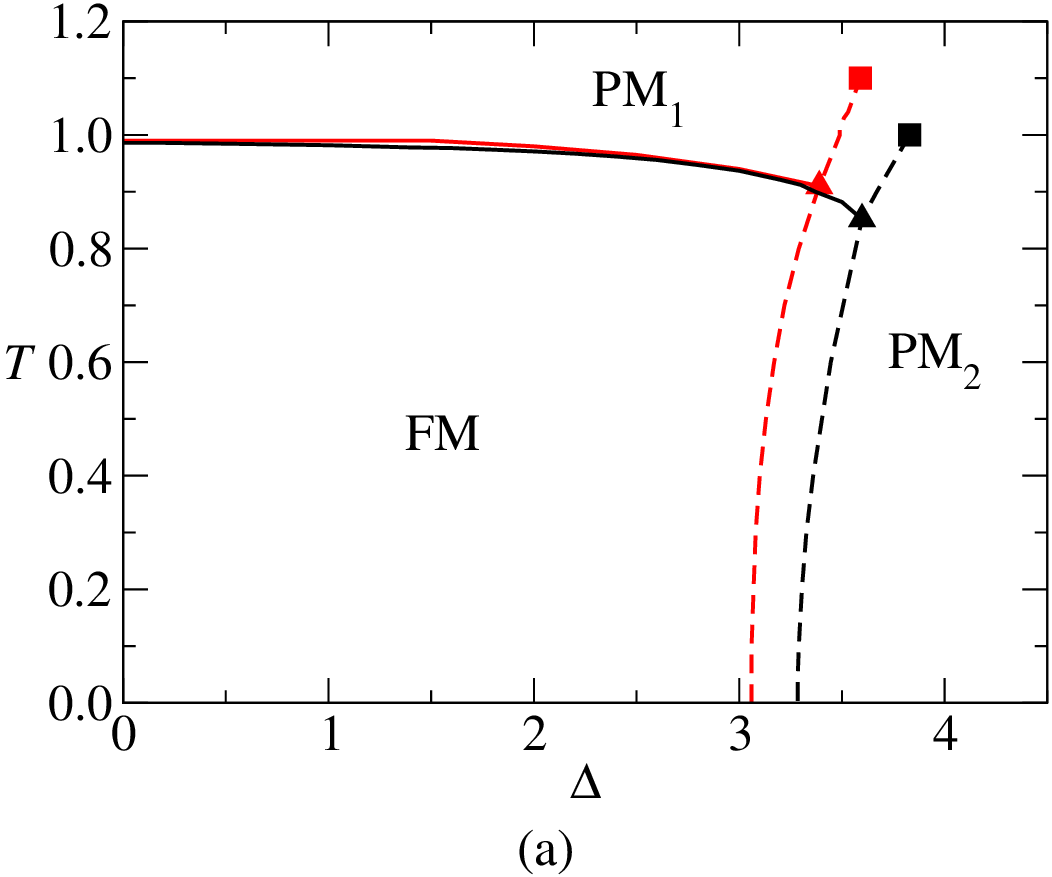}\hspace{0.5cm}
    \includegraphics[width=6.5cm,clip]{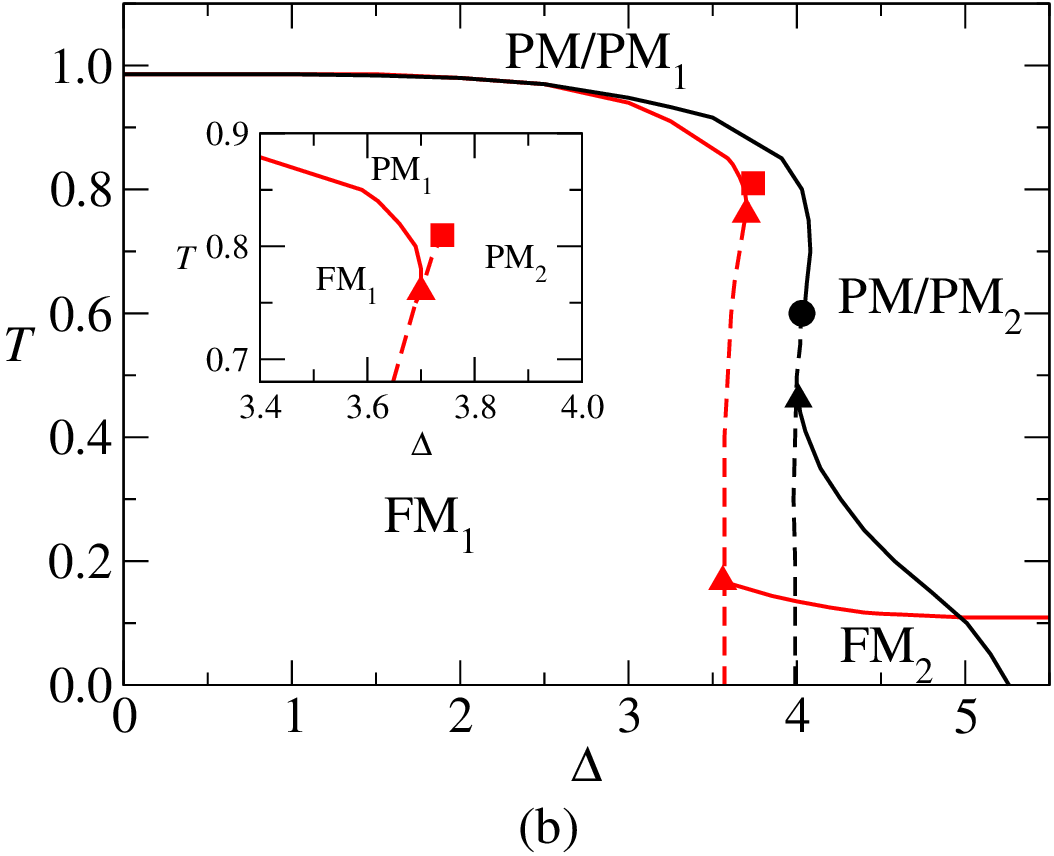}
    
    \vspace{0.5cm}
    \includegraphics[width=6.5cm,clip]{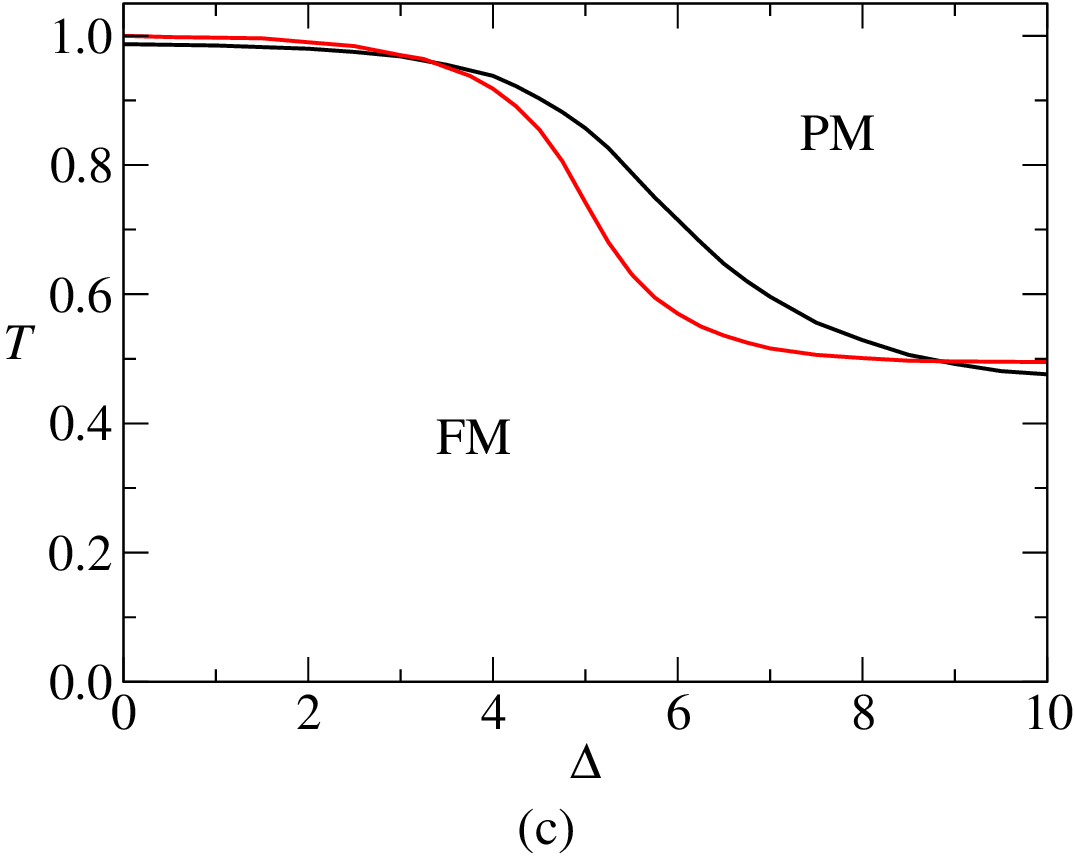}\hspace{0.5cm}
    \includegraphics[width=6.5cm,clip]{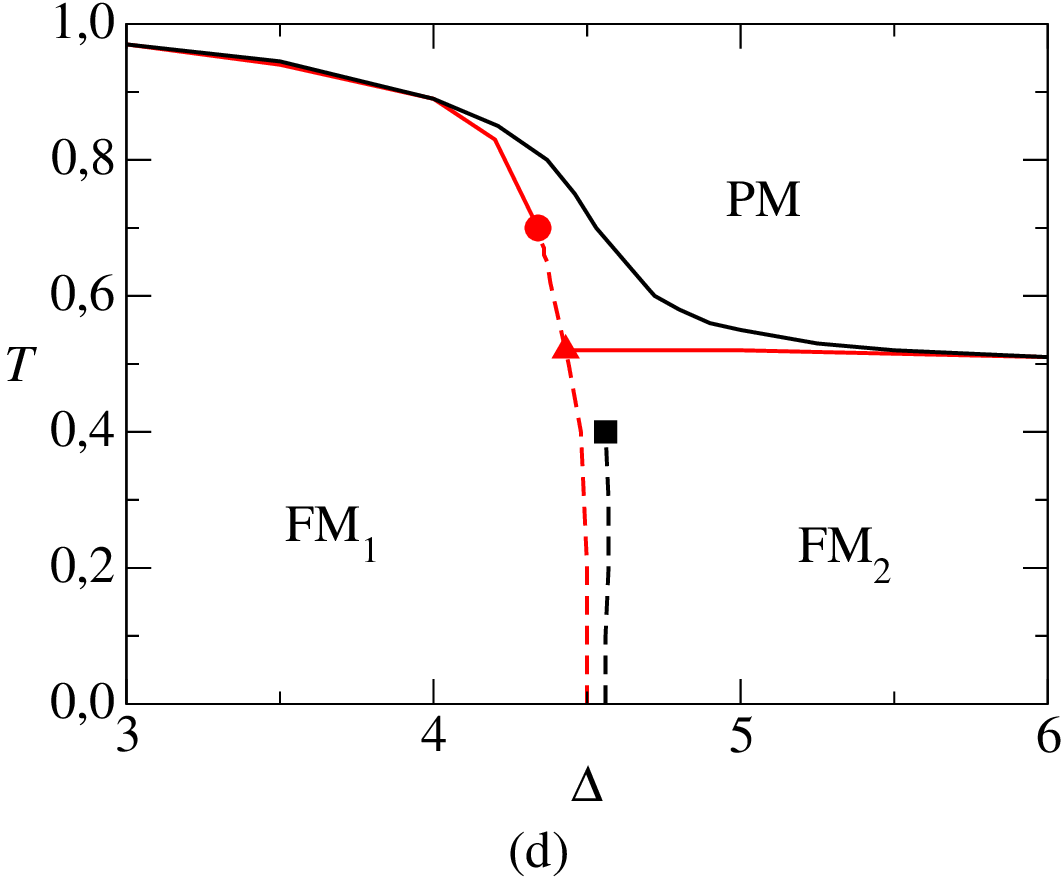}
    \caption{(a) Thermodynamic phase diagrams $T$ versus $\Delta$ for
      $p=1$ $K=5$, $c=4$ (black), $c=8$ (red). (b) The same, but for
      $p=0.85$; the inset shows in detail the vicinity of the critical
      points. (c) The same, but for $p=0.5$. (d) Thermodynamic phase
      diagrams for $p=0.5$, $K=5$, $c=25$ (black), $c=100$
      (red). Solid (dashed) lines correspond to continuous
      (discontinuous) transitions. Circles, squares and triangles
      represent tri-critical points, critical points and critical end
      points, respectively.  }
\label{fig:TD_5}
\end{figure}

To describe the finite temperature behavior, $T$ versus $\Delta$ phase
diagrams for $K=2$ and $K=5$ are presented in Figs. \ref{fig:TD_2} and
\ref{fig:TD_5}, respectively. For each $K$ value results for
representative disorder parameters $p=1$, $p=0.85$ and $p=0.5$, as
well as connectivity values $c=4$ and $c=8$, are shown. Results for
$p=0.5$ with $c=25$ and $c=100$ were also included, allowing for a
better comprehension of the convergence to the mean field approach,
which is expected for large $c$ (see ref. \cite{PhysRevB.60.1033}).

Smaller $c$ values, like $0<c<1$ are below the percolation limit $c=1$
preventing, thus, the appearing of ordered phases. This way, the
solutions would be $m=0$, $Q>0$. The most interesting feature is the
appearing of two paramagnetic phases, PM$_1$ and PM$_2$ (to be defined
below), depending on parameters $T$ and $\Delta$.

The ordered case, $p=1$ is shown in Figs. \ref{fig:TD_2}(a), for $K=2$
and \ref{fig:TD_5}(a), for $K=5$. If $K=2$, there is a FM phase at low
$T$, low $\Delta$ and a single PM phase elsewhere, with a continuous
transition at high temperature, a re-entrant discontinuous transition
at high $\Delta$ and a tricritical point (TCP) between them. TCPs,
critical points (CPs) and critical end points (CEPs) are indicated as
circles, squares and triangles in the figures. The re-entrant behavior
is illustrated in Fig. \ref{fig:mQf_T_8_.05}, described above. If
$K=5$, in addition to the FM phase there are two paramagnetic phases,
PM$_1$ and PM$_2$. The co-existence of PM$_1$ and PM$_2$ is typical of
models with a non-magnetic state $\sigma=0$, in which a sufficiently
large crystal field suppresses the active $\sigma=\pm 1$ states. The
high $Q$ and low $Q$ PM phases are named PM$_1$ and PM$_2$,
respectively. The transition from FM to PM$_1$ is continuous, while
the transition from PM$_2$ to FM and to PM$_1$ is discontinuous and
re-entrant, with a CEP where the two lines meet. The PM$_1$ - PM$_2$
discontinuous transition ends at CP.  The $p=1$ diagrams are similar
to those concerning the Bethe lattice approach reported in
\cite{Akheyan_1996}, although the re-entrant behavior in the
discontinuous transition is more pronounced in the present paper. The
re-entrant behavior in the ordered system with $K=2$ was also reported
in \cite{PhysRevB.60.1033}. As a further remark, our results are
qualitatively equivalent for both $c=4$ and $c=8$, although a lowering
$c$ appear to favour ordered phases.

Disorder, even for a moderate amount, i.e. $p=0.85$, unfolds the
ferromagnetic phase in two, namely FM$_1$ and FM$_2$. The first one is
reminiscent of the ordered system's FM phase. The second one, located
at low $T$ and large $\Delta$, arises consequently to disorder that
turns off the crystal field in a fraction $1-p$ of sites favouring the
active states in these sites, as stated above. Connectivity effects
become relevant. Figures \ref{fig:TD_2}(b) and \ref{fig:TD_5}(b) show
that, for $c=8$, FM$_2$ extends unbounded in $\Delta$, in contrast to
$c=4$, where there appears a zero temperature PM phase. We argue that
a moderate level of disorder is not a sufficient condition to
stabilize a FM$_2$ phase at large $\Delta$. Instead, it must be
associated to a large cooperative FM neighborhood. This condition is
found for $c=8$, but it is not for $c=4$. The random network with
$c=8$ and a moderate amount of disorder behaves similarly to a fully
connected one, whose mean-field results are reported in
\cite{PhysRevB.60.1033}. In both models there is a part of the FM$_1$
- PM$_2$ that is discontinuous. In our case, although the finite
connectivity, the random graph architecture still preserves a high
dimensional nature. Conversely, renormalization group results for
bi-dimensional systems, also reported in \cite{PhysRevB.60.1033}, show
that this transition is entirely continuous. To end this part,
additional qualitative differences between $K=2$ and $K=5$ for
$p=0.85$ should be reported. For $K=2$, $c=4$, there is a
discontinuous FM$_1$ - FM$_2$ transition that ends in a CP, shown in
the inset of Fig. \ref{fig:TD_2}(b). This way, the transition between
the two FM phases and the PM is always continuous and re-entrant, as
illustrated in Fig. \ref{fig:mQ_2_.85} . Conversely, for $K=5$ and
$c=4$ there is a CEP and a TCP in the FM - PM transition, as shows
Fig. \ref{fig:TD_5}(b). This figure also shows, detailed in the inset,
for $c=8$, a discontinuous PM$_1$ - PM$_2$ transition ending in a CP.

The scenario for a larger disorder, e.g. $p=0.5$, is shown in
Figs. \ref{fig:TD_2}(c) and \ref{fig:TD_5}(c) corresponding to $K=2$
and $K=5$, respectively. There is little to remark in these figures
beyond the $\Delta$-dependent continuous FM - PM transition. The
expectation for lower $p$ values is that the critical temperature
approaches a constant $T\sim 1$ for all $\Delta$. This behavior is
significantly distinct from the mean-field description for high
disorder \cite{PhysRevB.60.1033}. To investigate the behavior of the
highly disordered random network as $c$ increases, the phase diagrams
for $c=25$ and $c=100$, $K=2$ and $K=5$ were drawn, for $p=0.85$. The
results are shown in Figs. \ref{fig:TD_2}(d), for $K=2$ and
\ref{fig:TD_5}(d), for $K=5$. The results show that the convergence to
the fully connected scenario is faster for $K=5$. For $c=25$ the FM
phase unfolds in FM$_1$ and FM$_2$ with a discontinuous transition
between them ending in a CP. The fully connected scenario is observed
for $c=100$, with a CEP, a TCP and discontinuous FM$_1$ - PM
transition between them.

\section{Conclusions\label{sec:ccl}}

The BEG model with a disordered random crystal field was revisited, in
a random graph topology, employing a finite connectivity
technique. The disorder was introduced in the crystal field, as in
\cite{PhysRevB.60.1033} and through the random graph architecture. We
argue that, instead the crystal field, disorder could be introduced in
the bi-quadratic coupling constant and it would play a similar
role. Our model for disorder `turns off' the crystal field in a
fraction of sites, allowing to this fraction to assume active states
$\sigma=\pm 1$ without energetic penalty, even for large crystal field
values.

Models with an inactive state $\sigma=0$, like the ordered BEG model,
unfolds the PM phase in a high temperature PM$_1$ and a low
temperature PM$_2$. The main role that the disorder plays is to unfold
the FM phase in a high magnetization FM$_1$ and a low magnetization
FM$_2$. The last one survives at high crystal field values because the
crystal field is `turned off' in a finite fraction of sites. We fixed
$K=2$ and $K=5$.  Anti-ferromagnetic coupling constant $K<0$, as
reported in \cite{Kple2021} for the Bethe lattice with fixed
coordination number, allow for a richer thermodynamic scenario with
the appearing of a quadrupolar staggered phase. To do the same in a
random network architecture would require the introduction of a
sub-network or a random network of clusters, and this remains in our
scope for future works.

To end this work we resume the most relevant results.  i) We found
that the moderate disorder regime, e.g. $p=0.85$, is the more
sensitive to changes in the average connectivity, because the
stabilization of FM$_2$ relies on the cooperative effect of a large
neighborhood. Otherwise, for small $c$, a PM phase sets at low $T$ and
large $\Delta$. This is the regime where the finite connectivity
network becomes the more distinct from the fully connected one. ii)
For a large disorder, like $p=0.5$, the FM$_1$ - FM$_2$ discontinuous
transition and the associated CP disappear at low $c$ values, like
$c=4$ and $c=8$, appearing for $c$ as large as $c=25$. iii) A phase
diagram similar to the fully connected mean field description one only
appears for $c=100$ and $K=5$, but not for $c=100$ and $K=2$. This
suggests, in general lines, that some of the features observed in mean
field phase diagrams are artifacts that does not exist in most of the
real, finite connectivity physical systems.

\section*{Acknowledgements}

The authors thanks to Dr. Nilton Branco for fruitful discussions and
for carefully reading the manuscript. This work was supported, in
part, by CNPq (Conselho Nacional de Desenvolvimento Cient\'{\i}fico e
Tecnol\'ogico, Brazil).

\section*{Appendix: self consistent equation for the field distribution}

The site spin variables appearing in the inner exponential of the
replicated partition function, Eq. (\ref{part1}), are removed using
the identity
\begin{align}
1=\prod_{\alpha=1}^{n}\sum_{\sigma_{\alpha}}\delta_{\sigma_{\alpha}\sigma_{\alpha
    i}}=\sum_{\boldsymbol{\sigma}}\delta_{\boldsymbol{\sigma}\boldsymbol{\sigma}_{i}}\,,
\end{align}
where $\boldsymbol{\sigma}=\{\sigma_{1}\dots\sigma_{n}\}$ is a vector
of replicated spin variables and $\boldsymbol{\sigma}_{i}$ is the
replicated spin variable associated to spin $i$. Introducing the order
functions $P(\boldsymbol{\sigma})$ through the identity
\begin{align}
1 = \int\prod_{\boldsymbol{\sigma}}dP(\boldsymbol{\sigma})
\delta\Big[P(\boldsymbol{\sigma})-\frac{1}{N}
  \sum_{i}\delta_{\boldsymbol{\sigma}\boldsymbol{\sigma}_{i}}\Big]\,,
\end{align}
Eq. (\ref{part1}) becomes
\begin{align}
\langle Z^{n}
\rangle=\sum_{\boldsymbol{\sigma}_{1}\dots\boldsymbol{\sigma}_{n}}
\int\prod_{\boldsymbol{\sigma}}dP(\boldsymbol{\sigma})
d\hat{P}(\boldsymbol{\sigma}) &
\exp\Big\{\sum_{\boldsymbol{\sigma}}\hat{P}(\boldsymbol{\sigma})
P(\boldsymbol{\sigma})\\ +\frac{cN}{2}
\sum_{\boldsymbol{\sigma}\boldsymbol{\sigma}'} P(\boldsymbol{\sigma})
& P(\boldsymbol{\sigma}')\Big(\mathrm{e}^{\frac{\beta J}{c}
  \sum_{\alpha} \sigma_{\alpha}\sigma_{\alpha}^{\prime}+\frac{\beta
    K}{c}\sum_{\alpha} \sigma_{\alpha}^{2}\sigma_{\alpha}^{\prime
    2}}-1\Big) \nonumber \\ & -
\frac{1}{N}\sum_{\boldsymbol{\sigma}}\hat{P}(\boldsymbol{\sigma})
\sum_{i}\delta_{\boldsymbol{\sigma}\boldsymbol{\sigma}_{i}}
\Big\}\langle\mathrm{e}^{-\beta\sum_{\alpha
    i}\Delta_{i}\sigma_{i\alpha}^{2}}\rangle_{\boldsymbol{\Delta}} \,.
\nonumber
\end{align}
Summing over the spin variables $\boldsymbol{\sigma}_{i}$ and changing
variables $\hat{P}(\boldsymbol{\sigma})\rightarrow
N\hat{P}(\boldsymbol{\sigma})$, Eq. (\ref{RSsp}) is
obtained. Expanding the exponential en Eq. (\ref{RS1}) and inserting
the RS Ansatz, we obtain
\begin{align}
P(\boldsymbol{\sigma})=&\sum_{k=0}^{\infty} P_k
\Big\langle\mathrm{e}^{-\beta\Delta\sum_{\alpha}\sigma_{\alpha}^{2}}
\Big\rangle_{\Delta}\int\prod_{l=1}^{k}\frac{\mathcal{D}W(x_l{,}y_l)}
           {\Big(\sum_\sigma\mathrm{e}^{\beta x_l\sigma_l + \beta
               y_l\sigma^{2}_l}\Big)^n}
           \exp\sum_{\alpha=1}^n\ln\chi_{\sigma_{\alpha}}(x_l{,}y_l)\,.
\end{align}
Now we withdraw the $\sigma_{\alpha}$ variables outside of the $\log$
using the identity $\sum_{\sigma}\delta_{\sigma\sigma_{\alpha}}=1$,
\begin{align}
\sum_{\alpha=1}^{n}\log\chi_{\sigma_{\alpha}}(x_{l},y_{l}) =
\sum_{\alpha=1}^{n}\sum_{\sigma}\delta_{\sigma\sigma_{\alpha}}
\ln\chi_{\sigma}(x_{l},y_{l})\,.
\end{align}
The Kronecker delta representation for the spin states
$\sigma=\{-1,0,1\}$ is given by
\begin{align}
\delta_{\sigma\sigma_{\alpha}} = 1 - \sigma^{2} - \sigma_{\alpha}^{2}
+ \frac{1}{2}\sigma\sigma_{\alpha} +
\frac{3}{2}\sigma^{2}\sigma_{\alpha}^{2}\,.
\end{align}
Summing over $\sigma$ we get, after some algebra,
\begin{align}
\nonumber P(\boldsymbol{\sigma})=\sum_{k=0}^{\infty} P_k
\Big\langle\int\prod_{l=1}^{k} & \frac{\mathcal{D}W(x_{l}{,}
  y_{l})}{\Big(\sum_{\sigma}\mathrm{e}^{\beta x_{l}\sigma_{l} + \beta
    y_{l}\sigma^{2}_{l}}\Big)^{n}}
\exp\Big\{\Big(\sum_{\alpha}\sigma_{\alpha}\Big)\sum_{l=1}^{k}\phi(x_{l},y_{l})
\\ & +
\Big(\sum_{\alpha}\sigma^{2}_{\alpha}\Big)\sum_{l=1}^{k}\psi(x_{l},y_{l})
-\Big(\sum_{\alpha}\sigma^{2}_{\alpha}\Big)\beta\Delta\Big\}
\Big\rangle_{\Delta} \,.
\end{align}
Substitution of RS Ansatz in the LHS and taking the limit
$n\rightarrow 0$
\begin{align}
\int \mathcal{D}& W(x{,}y)\mathrm{e}^{\beta
  x\sum_{\alpha}\sigma_{\alpha} + \beta
  y\sum_{\alpha}\sigma^{2}_{\alpha}} = \int dx dy
\Big\{\sum_{k=0}^{\infty} P_k \Big\langle\int\prod_{l=1}^{k}
\mathcal{D}W(x_{l},y_{l})\\ \nonumber &\times\delta\Big[x
  -\beta^{-1}\sum_{l}\phi(x_l{,}y_l)\Big]\delta\Big[y+
  \Delta-\beta^{-1}\sum_l\psi(x_l{,}y_l)\Big]\Big\rangle_{\Delta}\Big\}
\mathrm{e}^{\beta x\sum_{\alpha}\sigma_{\alpha} + \beta
  y\sum_{\alpha}\sigma^{2}_{\alpha}}\,.
\end{align}
Comparing both sides of this equation we obtain Eq. (\ref{fieldist}).

\bibliography{file2}

\end{document}